\documentclass[12pt,preprint]{aastex}
\received{99/99/9999}
\revised{99/99/9999}
\accepted{99/99/9999}
\ccc{code}
\cpright{type}{yyyy}
\slugcomment{Variety of environment of a GRB progenitor }
\shorttitle{Variety of environment of a GRB progenitor }
\shortauthors{Yonetoku et al.}

\begin{document}

\title{ NON-EQUILIBRIUM IONIZATION STATES OF GRB ENVIRONMENTS }
\author{Daisuke Yonetoku\altaffilmark{1,2} and 
Toshio Murakami\altaffilmark{1,2}}
\email{yonetoku@astro.isas.ac.jp, murakami@astro.isas.ac.jp}
\author{Kuniaki Masai\altaffilmark{3},
Atsumasa Yoshida\altaffilmark{4},
Nobuyuki Kawai\altaffilmark{2},
Masaaki Namiki\altaffilmark{5}}

\altaffiltext{1}{Institute of Space and Astronautical Science, 3-1-1, 
Yoshinodai, Sagamihara, Kanagawa 229-8510, Japan}
\altaffiltext{2}{Department of Physics, Tokyo Institute of Technology, 
2-12-1 Ookayama, Meguro, Tokyo 152-0033, Japan}
\altaffiltext{3}{Department of Physics, Tokyo Metropolitan
University, 1-1 Minamiosawa, Hachioji, Tokyo 192-0397, Japan}
\altaffiltext{4}{Department of Physics, Aoyama Gakuin University,
6-16-1, Chitosedai, Setagaya, Tokyo, 157-8572  Japan}
\altaffiltext{5}{The Institute of Physical and Chemical 
Research, 2-1, Hirosawa, Wako 351-0198, Japan}

\begin{abstract}
  Iron spectral features are thought to be the best tracer of a
progenitor of gamma-ray bursts (GRBs). The detections of spectral
features such as an iron line and/or a Radiative Recombination edge
and Continuum (RRC) were reported in four X-ray afterglows of
GRBs. However their properties were different each other burst by
burst. For example, Chandra observation of GRB~991216 reported both
the strong H-like iron line together with its RRC. On the contrary,
Yoshida et al. (2001) report only a detection of the strong RRC in
GRB~970828 with ASCA.  Since it is difficult to produce the strong
RRC, we have to consider special condition for the line and/or the RRC
forming region. In this paper, we point out a possibility of a {\it
``non-equilibrium ionization state''} for the line and the RRC forming
region.
\end{abstract}


\section{Introduction}
 Gamma-Ray Bursts (GRBs) are remote at cosmological distances and thus
their released energies in $\gamma$-ray photons are almost $\sim
10^{52}~{\rm erg}$ \cite{kul99}. Although fireball models
\cite{rees92, piran98} or a cannonball model \cite{dar00} can explain
many observational properties of GRBs, a progenitor of GRB is yet to
be solved.  An important key to probe a progenitor of GRB is the
detection of iron spectral features in X-ray afterglows. Iron is the
most abundant heavy element and fall in the middle of the observed
energy range of X-ray satellites.  Strong iron features can be
produced in a dense gas environment. Therefore, the detection is the
best evidence of a collapsing model of a massive star such as
Hypernova/Collapsar or Supranova \cite{pac98, woo93, vie98}.

 The iron features were already reported in four X-ray afterglows. The
redshifted iron emission line was discovered in the X-ray afterglow of
GRB~970508 with BeppoSAX \cite{piro99}, and the redshift was
consistent with the distance of a host galaxy.  An independent
discovery of a redshifted iron emission line was reported for
GRB~970828 with ASCA \cite{yoshi99}. Assuming a He-like K$\alpha$
line, the spectral feature was once interpreted as an emission line
with a redshift of $z = 0.33$.  However, the Keck observation of the
host galaxy, which was discovered later in 1998, revealed the redshift
of $z = 0.9578$ \cite{dj00}. The discrepancy in distance with ASCA and
moreover the temporal detections of both forced us to doubt the
reality of existence of the iron features before the confirmation by
the Chandra detection.

  In 1999, Chandra observed GRB~991216 at 37 hours after the GRB with
the ACIS-S/HETG instruments, and Piro et al. (2000) reported the
detection of both the iron emission line together with the Radiative
Recombination edge and Continuum (RRC) from fully ionized
iron with high statistical significance of $4.7~\sigma$ and
$3.0~\sigma$ respectively. The redshift of the line and the RRC were
consistent with the host galaxy, and the line was broaden to
$\sigma_{\rm line} = 0.23 \pm 0.07~{\rm keV}$ suggesting a moving
ejecta.  Soon after the Chandra detection, Antonelli et al. (2000)
also reported the probable iron emission line in the spectrum of
GRB~000214 with $\sim 3.2~\sigma$ confidence level. This was the
second detection of the iron with BeppoSAX. However the redshift of a
host galaxy of GRB~000214 is unknown. Therefore, it remains a question
whether the observed feature is the iron emission line or the RRC.

  Motivated by the RRC detection with Chandra, Yoshida et
al. \cite{yoshi01} apply the RRC for the ASCA data of GRB~970828 using
the redshift of $z = 0.9578$, and obtain the edge energy, which is
consistent with $9.28~{\rm keV}$. This possibility was earlier
suggested in the paper by Djorgovski et al. (2001). However, there was
no iron K$\alpha$ line detected at the expected energy from
GRB~970828. The upper limit of the line flux is $< 1.5 \times
10^{-6}~{\rm photons~cm^{-2}s^{-1}}$ or $120$ eV in equivalent width
(EW). We must explain the no detection of K$\alpha$ line.

 Related to the iron features, we should note the fact that the iron
features were found only in a small fraction ($< 10$ \%) of X-ray
afterglows, and sometimes the features were observed only during a
certain interval. BeppoSAX has tried to search lines in eleven X-ray
afterglow spectra, but found only one from GRB~970508 at the end of
1999 (Piro; private communication).  Most X-ray afterglows did show
only upper limits.  Especially, Yonetoku et al. (2000) set an
extremely low upper limit of almost $100$ eV in EW for the bright
GRB~990123 with ASCA. The published intensities of the observed iron
K$\alpha$ line and the RRC are summarized in table~\ref{tab:Fe}. There
is a wide variety in the iron emission and the RRC intensity. In this
paper, we try to explain these varieties of the iron features with the
assumption that the line-emitting plasma state is in ``{\rm
non-equilibrium ionization (NEI) state}'' which has a low electron
temperature as compared to the ionization degree.

\section{Spectral Simulation for Non-equilibrium Plasma}

  In paper-I, the observed flux of the RRC and the upper limit for the
iron K$\alpha$ line are given for GRB~970828 with $90$ \% confidence
errors.  So, we focus on the ratio of the integrated RRC flux to the
iron intensity: $F_{\rm RRC} / F_{\rm line} $, which is free from the
continuum.  The observed ratio of: $F_{\rm RRC} / F_{\rm line} > 3.3$
in $90$ \% statistical lower limit is very hard to reproduce. The case
of a strong RRC without K$\alpha$ line looks abnormal. In fact, BeppoSAX
detected the strong K$\alpha$ line and the Chandra observation showed
both features of H-like iron with the flux ratio: $F_{\rm RRC} /
F_{\rm line} \sim 1$.  Therefore, we are forced to consider a
different condition between ASCA and other results containing the
strong K$\alpha$ line. 

The no K$\alpha$ line of highly ionized iron accompanied 
by the recombination edge gives a constraint on the possible 
emission mechanisms.  If the line were produced due to excitation by
electron-impact, the RRC would be
hidden  by a much more intense thermal bremsstrahlung.
Also the synchrotron emission, which is thought to dominate the
afterglow, would be significantly overlaid by the thermal
bremsstrahlung.  Therefore, we need a condition that the iron is
highly ionized but the involved electron energy is low. This is a 
radiative recombination in the NEI ($T_{e} < T_{z}$) state or 
charge exchange must be responsible for the iron emission.

 In this section, we show spectra by numerical calculations in the NEI
plasma state, not depending on the specific model.  To explain the
observed strong RRC without the K$\alpha$ line of iron, we calculate
the emissivity using the NEI plasma radiation code \cite{masa94}. The
code employs three mechanisms for the continuum, such as free-free
emission, two-photon decay and radiative recombination. For line
emissions, as well as excitation by electron-impact, fluorescence lines 
due to ionization and cascade lines due to recombination are
taken into account.  Radiation properties of a plasma were described by 
two parameters of the electron temperature ($T_{e}$) and the ionization 
degree represented in units of temperature ($T_{z}$), assuming a cosmical 
abundance. We study the emissivities in the range of 
$0.1 < kT_{e} < 10~{\rm keV}$ and $0.1 < kT_{z} < 100~{\rm keV}$ 
in every $0.1~{\rm keV}$ step.  We show representative spectra 
in figure~\ref{fig:spec_sim}.

 The strong RRC compared with the K$\alpha$ line can be formed only in 
the regime of $T_{e} < T_{z}$, recombining plasma condition.
This condition is, however, attained by several situations as discussed
later. We intend to find the condition of the plasma to
account for the observed flux ratio ($F_{\rm RRC}/F_{\rm line}$),
which is free from specific modelling with iron abundance, the
emission measure, the geometry and so forth if the line and RRC
are emitted from the same cite. Thus, for a given $T_{e}$ and $T_{z}$ 
a priori, we carried out calculations of the emissivity ratio in 
the above wide range of $T_{e}$ -- $T_{z}$ space.

 Figure~\ref{fig:spec_sim}-c and \ref{fig:spec_sim}-d are the
simulated plasma emissivities with a cosmical abundance to explain the
ratio: $F_{\rm RRC} / F_{\rm line}$ of GRB~991216 and GRB~970828
respectively.  The ratios of an integrated RRC flux to K$\alpha$ lines
of the simulated ionization state are $F_{\rm RRC}/F_{\rm line} \sim
1$ and $\sim 4$ for figure~\ref{fig:spec_sim}-c and
\ref{fig:spec_sim}-d respectively, and the ratios are within the
observational values in $90$ \% statistical error.  In the
calculation of the ratios, the line components (mostly the blend of Fe
and Ni) very close to the edge of the RRC are included to the RRC
component, due to the limited energy resolving power of the SIS
detectors onboard ASCA.

\section{The Reason for Strong RRC and Weak K$\alpha$ Line}

 To produce the strong RRC in quantum number $n = 1$, the iron must be
almost fully ionized. The H-like K$\alpha$ line, which was
observed with Chandra, dominates other ionization states at a
temperature of $kT_{z} > 20~{\rm keV}$. Above the temperature, Fe
XXVII consists of more than 70 \% of iron, thus we assume $kT_{z} >
20~{\rm keV}$ in the following discussion.  In a condition of high
electron temperature of $kT_{e} \sim kT_{z} > 20~{\rm keV}$, i.e.,
equilibrium ionization, the capture rate of free electrons is small
and the emissivity of the line and RRC also becomes small. Moreover,
the free-free emission from high $T_{e}$ electrons dominates at the
hard X-ray band. Therefore, the RRC may not be observed because it can
be obscured by the free-free component.

The cross section of the electron capture into $n$ th quantum state can 
be expressed as
\begin{eqnarray}
\sigma_{n} \propto \frac{1}{n^{3}} \bigl(  \frac{3}{2} \frac{kT_{e}}
{\epsilon}
+ \frac{1}{n^{2}}\bigr)^{-1}
\end{eqnarray}
where $\epsilon$ is an ionization energy of $9.28~{\rm keV}$ for H-like 
iron \cite{naka01}.
Thus, the best condition to form the strong RRC would be the case of 
$k T_{e} \sim \epsilon \, (\ll k T_{z})$.  In such a plasma state, 
$\sigma_{n} \propto n^{-3}$ and then most of free electrons
recombine directly into the ground state ($n = 1$), compared to the 
$n \ge 2$ levels.  However, free-free emission dominates the continuum.  

With decreasing $k T_{e}$, the recombination rate increases, while 
free-free emission becomes suppressed.  Recombination into
$n \ge 2$ increases relatively and produces line emission.  Thus, the 
K$\alpha$ line can be enhanced by cascades from $n \ge 3$ excited 
levels.  This is the case for He-like K$\alpha$, but H-like K$\alpha$ 
(Ly$\alpha$) is little affected; a considerable fraction comes to 
direct transition to the ground state.

  We summarize the above discussions in view point of the intensity of
the RRC and K$\alpha$ line. The plasma state with the strong RRC but
the weak K$\alpha$ line, which was observed with ASCA, is realized
when $kT_{e}$ is slightly less than $\epsilon$ but in high $kT_{z}$.
The ratio of the RRC to K$\alpha$ lines ($F_{\rm RRC}/F_{\rm line}$)
from the numerical calculations is shown in figure~\ref{fig:Ratio} as
the function of $T_{e}$ and $T_{z}$. The peak of the ratio appears at
around $kT_{e} = 4~{\rm keV}$ and $kT_{z} = 100~{\rm keV}$ in the
calculated range. The condition of $F_{\rm RRC}/F_{\rm line} > 3.3$ of
GRB~970828 can be explained by this result.

\section{Discussion}

  A high ionization degree of $kT_{z} \sim 100~{\rm keV}$ and low
electron temperature of $kT_{e} \sim 1~{\rm keV}$ are required when we
reproduce the high ratio of the RRC to the iron line more than
$3.3$. This condition may be attained in the situations: (i)
photoionizations by X-rays or (ii) rapid cooling due to rarefaction.

In the case (i), also a fluorescence K$\alpha$ line of energies 6.4 --
6.5 keV of partially ionized iron is likely accompanied.  Especially, 
the photoionizations of a neutral circumstellar gas by the
initial bright flash of GRB, iron lines in the low ionization states
are expected but not observed. The line observed with Chandra was
purely H-like (Piro et al. 2000).  Therefore, the iron atoms should be
fully ionized by the time of the observed iron emission.  However, if 
the line and the RRC emitting region is illuminated continuously by a 
hidden intense beam, discussed by Rees et al. (2000), it can achieve 
the NEI ($T_{e} < T_{z}$) state by the photoionization process. 
Even if this process realize, the mean energy of the hidden-beam 
photons which illuminate the line emitting region should not be largely 
greater than the edge energy of $9.28~{\rm keV}$, since the observed 
values of $T_{e}$ were low of $kT_{e} = 0.8^{+1.0}_{-0.2}~{\rm keV}$ at 
the rest frame for ASCA and $k T_{e} \gtrsim 1~{\rm keV}$ for 
Chandra respectively.

  We pay attention to the case (ii) of a rapid adiabatic expansion of 
a highly ionized hot plasma.  This sort of mechanism has already been 
investigated by Itoh \& Masai (1989) for a supernova which explodes 
in the circumstellar matter ejected during its progenitor's supergiant 
phase. They show that when the blast shock breaks out of the dense 
circumstellar matter into a low-density interstellar medium, a 
rarefaction wave propagates inward into the shocked hot plasma. Then the 
hot plasma expands adiabatically and loses its internal energy quickly.  
The electron temperature $T_e$ decreases, but the ionization degree $T_{z}$
temporally remains high, because the recombination time scale becomes 
much longer due to the low density.  The mean temperature of 
the shocked matter, $kT \sim 100~(v_{s}/10^9~{\rm cm~s^{-1}})^2~{\rm keV}$
where $v_{s}$ is the shock velocity, drops by about two orders of 
magnitude for the density contrast (ratio) of the dense matter to 
the ambient medium of $\sim 10^{3}$ in their hydrodynamic calculations.
Therefore, the plasma can achieve the NEI 
($T_{e} \sim 1~{\rm keV}, T_{z} \sim 100~{\rm keV}$) state naturally,
and emit the strong RRC with the weak K$\alpha$ line of 
$F_{\rm RRC}/F_{\rm line} \sim 4$.

 If these mechanisms work, we can estimate an emission measure for
GRB~970828 using the observed photon flux of the RRC 
($\mathcal{F}_{\rm RRC}$), the distance $D$ and the integrated 
emissivity ($\varepsilon_{\rm RRC}$) shown in 
the figure~\ref{fig:Ratio} with the cosmical abundance,
\begin{eqnarray}
n^{2}V 
\sim 10^{68}
\bigl( \frac{D}{3~{\rm Gpc}} \bigr)^{2}
\bigl( \frac{\mathcal{F}_{\rm RRC}}{1.7 \times 10^{-5}} \bigr)
\bigl( \frac{\varepsilon_{\rm RRC}}{2 \times 10^{-16}} \bigr)^{-1}
\end{eqnarray}
Although these $n$ and $V$ are coupled with each other and 
highly depend upon the specific model, we may conclude
that the density of line emitting region is considerably high.

The RRC and/or the line of a large EW suggest a recombing ($T_{e} < T_{z}$) 
condition, which can be realized by the photoionization or rarefaction.
It should be noted that most of X-ray afterglow did not show both 
emission features; NEI is not always the case.

\section{Acknowledgment}
  We would like to thank Shri Kulkarni and George Djorgovski for their
comments and suggestions about the distance to the host galaxy before
the publication. This work was done under the support of
Grant-in-Aides for the Scientific Research (Nos. 12640302) by the
Ministry of Education, Culture, Sports, Science and Technology.

\clearpage
%figure 1 (simulated emissivities)
\figcaption[]{Simulated emissivities, convolved with the energy 
resolution of the ASCA-SIS for the cases; (a) $T_{z} = 10~{\rm keV}$, 
$T_{e} = 10~{\rm keV}$ (equilibrium), (b) $T_{z} = 1~{\rm keV}$, 
$T_{e} = 10~{\rm keV}$, (c) $T_{z} = 15~{\rm keV}$, $T_{e} = 2~{\rm keV}$ 
and (d) $T_{z} = 100~{\rm keV}$, $T_{e} = 1~{\rm keV}$. The solid
lines represent emissivity of continuum and the dotted ones are for 
emission lines. The figure 1-c and 1-d are simulated for representing 
the cases of GRB~991216 and GRB970828 respectively only in view point 
of the observed ratios of $F_{\rm RRC} / F_{\rm line}$. We do not intend 
to reproduce the spectral shapes, which mostly consist of a non-thermal
component.
\label{fig:spec_sim}
}

\clearpage
%figure 2 (RRC/Line Ratio)
\figcaption[]{Ratios of the integrated emissivity of the RRC structure 
to the K$\alpha$ lines: ($F_{\rm RRC} / F_{\rm line}$) for several 
ionization states in 3-dimensional axis. 
We calculate the ratio of the emissivity in each $T_{z} = 5~{\rm keV}$
and $T_{e} = 0.5~{\rm keV}$ steps. The straight line labeled
``equilibrium'' indicates the condition of $T_{e} = T_{z}$.  The
recombination process enhances only in the $T_{e} < T_{z}$
region. Within our calculated range, the observed ratio of $F_{\rm
RRC} / F_{\rm line} > 3.3$ for GRB~970828 can be attained in the
range of more than $T_{z} = 80~{\rm keV}$, limiting the $T_{e} =
1~{\rm keV}$, which is observed in the RRC structure.
\label{fig:Ratio} 
}

\clearpage
\begin{table}[htbp]
\begin{center}
\caption{Intensity of Iron Line and RRC} 
\label{tab:Fe}
\vspace{2mm} 
\begin{tabular}{c|c|c|c|c|c} \hline\hline
GRB & $z$  & {\bf $F_{\rm line}$} & {\bf $\sigma$} 
& {\bf $F_{\rm RRC}$} & {\bf $kT$} \\
Name&& (${\rm photons~cm^{-2}s^{-1}}$) & (keV) 
& (${\rm photons~cm^{-2}s^{-1}}$) & (keV) \\
\hline
$970508\hspace{1pt}^{\ast}$ & 0.835 & $(3.0 \pm 2.0) \times 10^{-5}$ 
& --- & --- & --- \\
$970828\hspace{1pt}^{\dagger}$ & 0.958 & $< 1.5 \times 10^{-6}$ & --- 
& $ 1.7^{+ 6.4}_{-1.2} \times 10^{-5}$ & $0.8^{+1.0}_{-0.2}$ \\
$991216\hspace{1pt}^{\sharp}$ & 1.020 & $(3.2 \pm 0.8) \times 10^{-5}$ 
& $0.23 \pm 0.07$ & $3.8 \pm 2.0$ & $>1.0$ \\
$000214\hspace{1pt}^{\natural}$ & ---   & $(9 \pm 3) \times 10^{-6}$ 
& --- & --- & --- \\
\hline\hline
$990123\hspace{1pt}^{\star}$ & 1.600 & $< 3.3 \times 10^{-6}$ 
& --- & --- & ---\\
$990704\hspace{1pt}^{\star}$ & ---   & $< 4.7 \times 10^{-6}$  
& --- & --- & --- \\
\hline
\end{tabular}
\end{center}
\end{table}
\vspace{-20pt}
\par\noindent
$^{\ast}$ Piro et al. 1999 ; the line intensity was variable.\\
$^{\dagger}$ Yoshida et al. 2001 ; the RRC was temporal. \\
$^{\sharp}$ Piro et al. 2000\\
$^{\natural}$ Antonelli et al. 2000\\
$^{\star}$ Yonetoku et al. 2000\\

\clearpage
\begin{figure}[htb] 
\begin{center} 
\rotatebox{270}{\scalebox{0.25}{\includegraphics{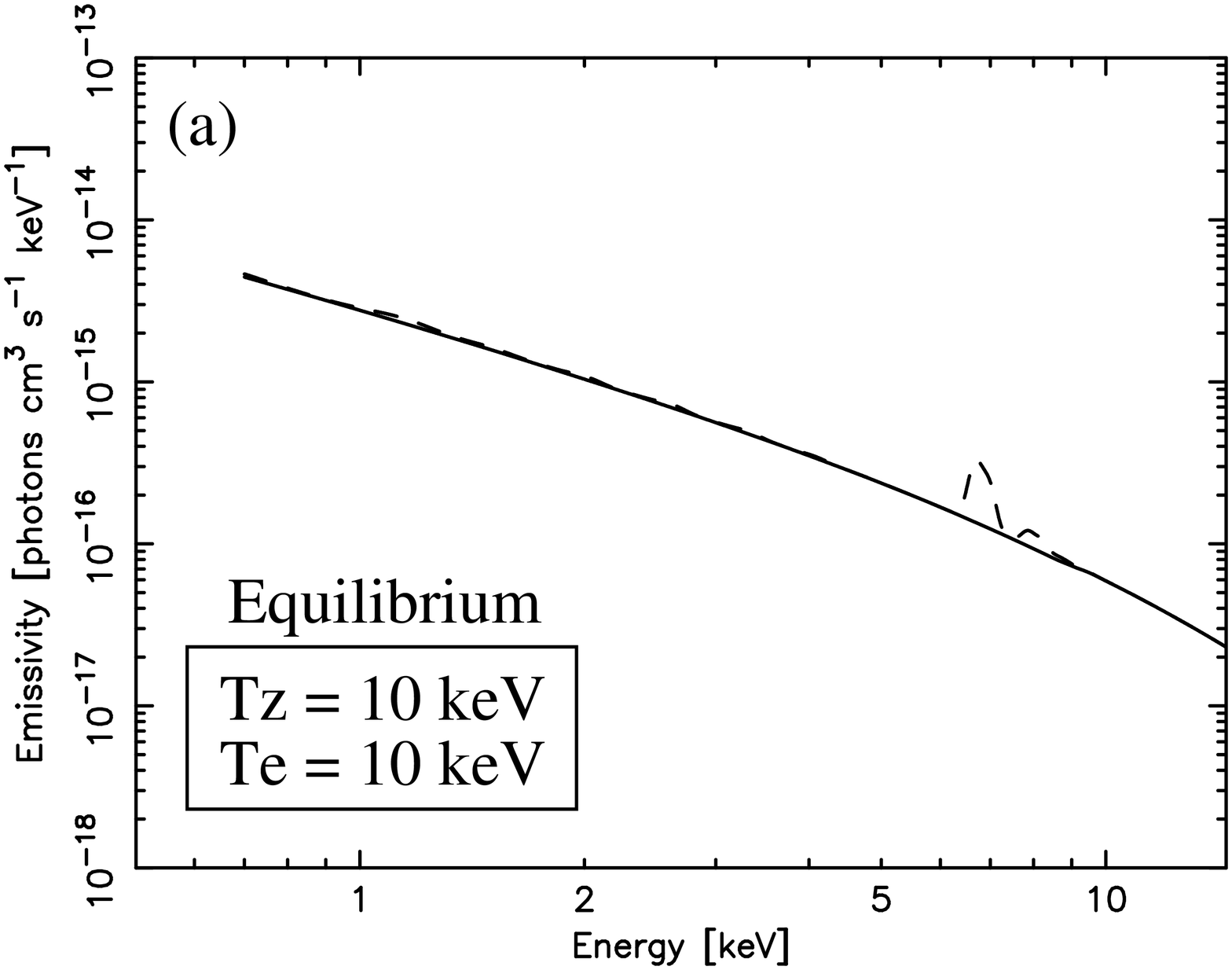}}} 
\rotatebox{270}{\scalebox{0.25}{\includegraphics{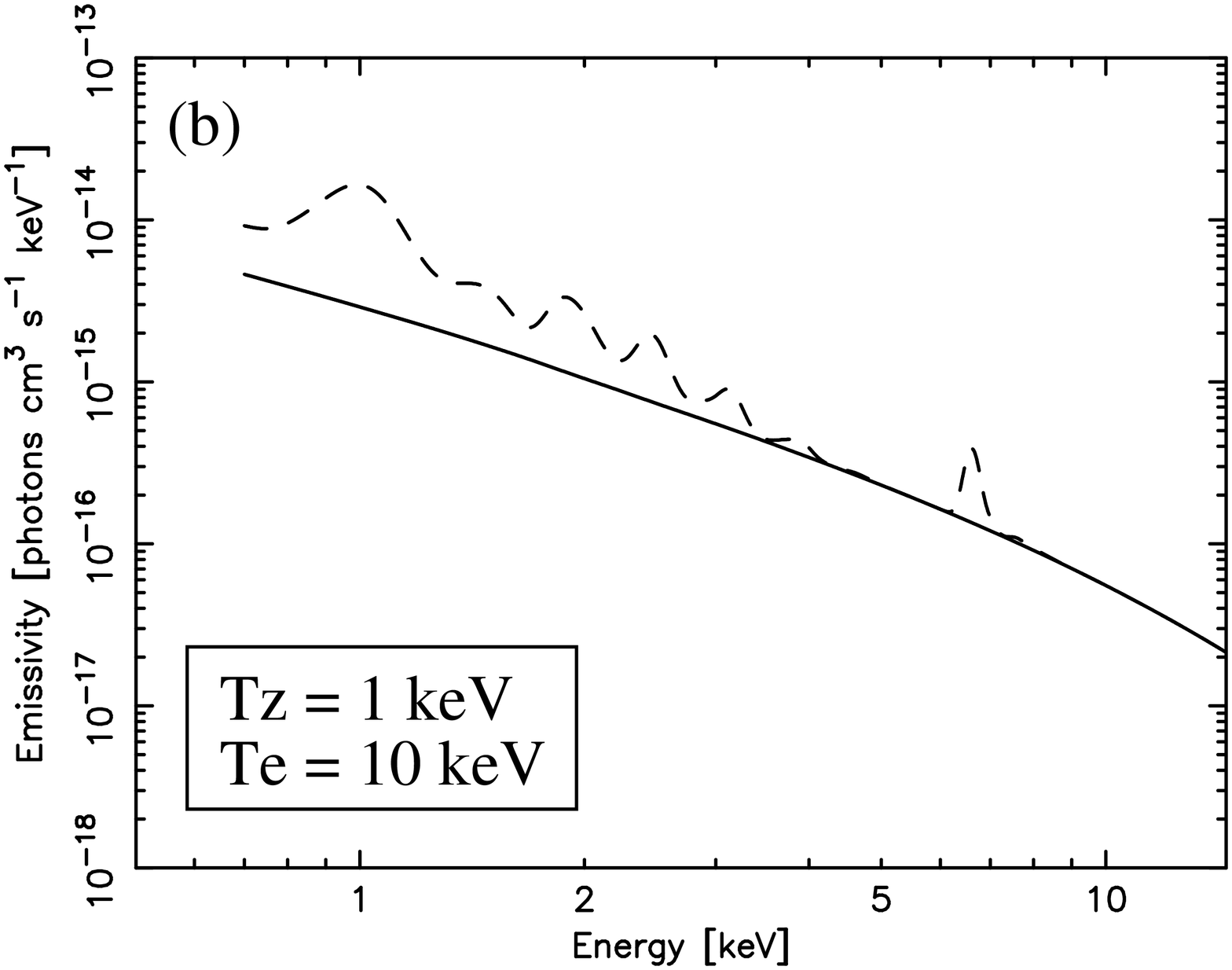}}} 
\rotatebox{270}{\scalebox{0.25}{\includegraphics{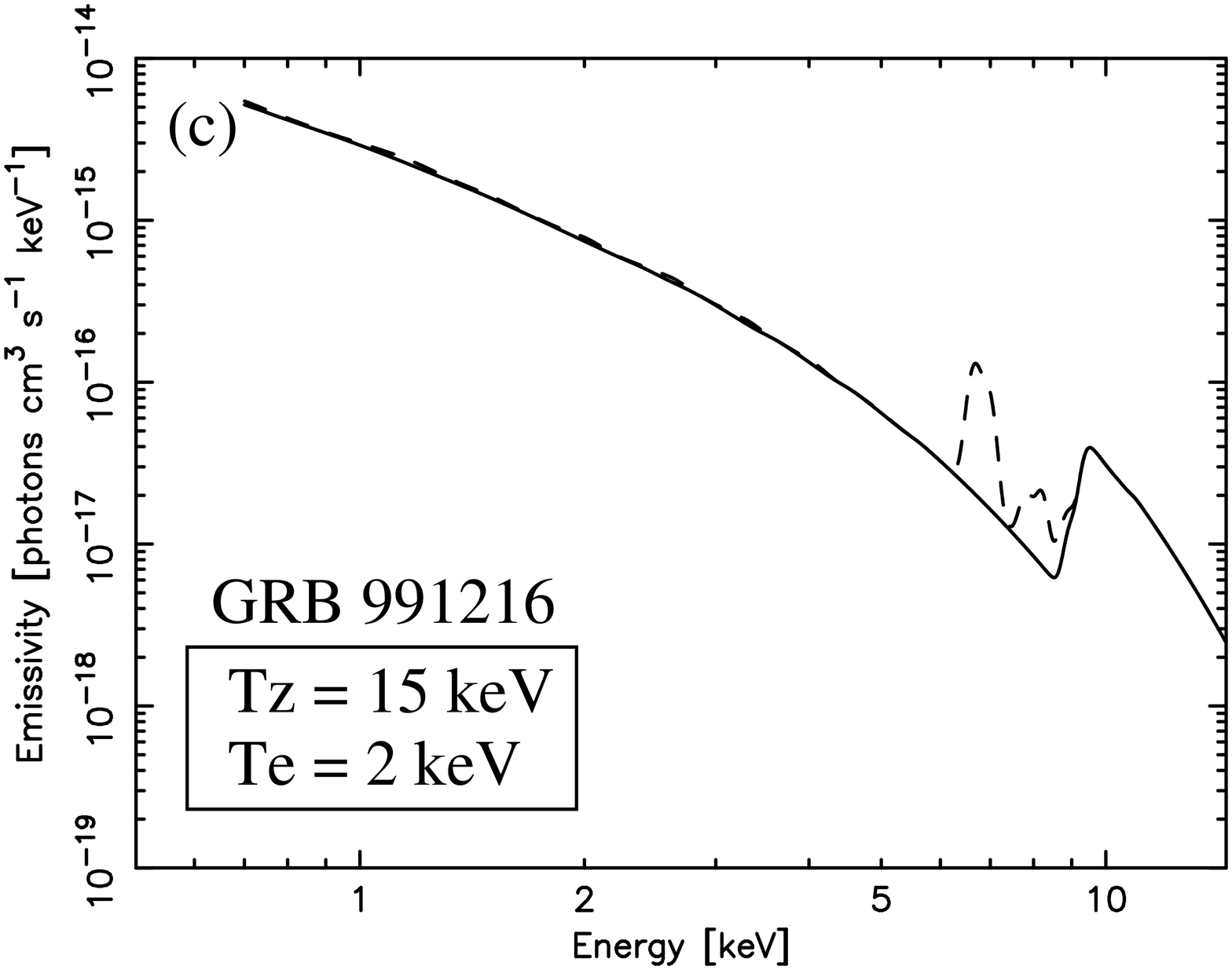}}} 
\rotatebox{270}{\scalebox{0.25}{\includegraphics{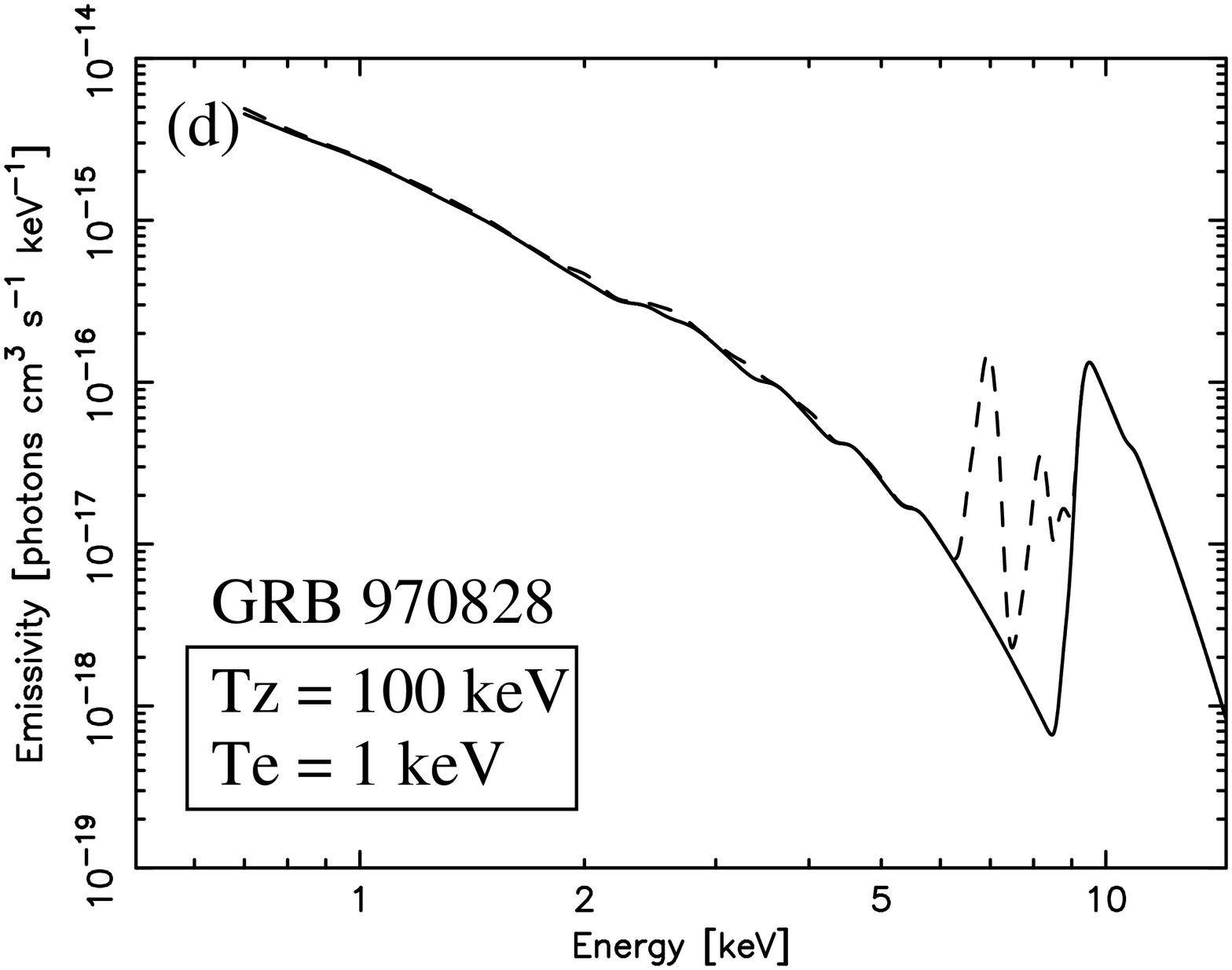}}} 
\end{center} 
\end{figure} 

\clearpage
\begin{figure}[htb] 
\begin{center} 
\rotatebox{0}{\scalebox{0.60}{\includegraphics{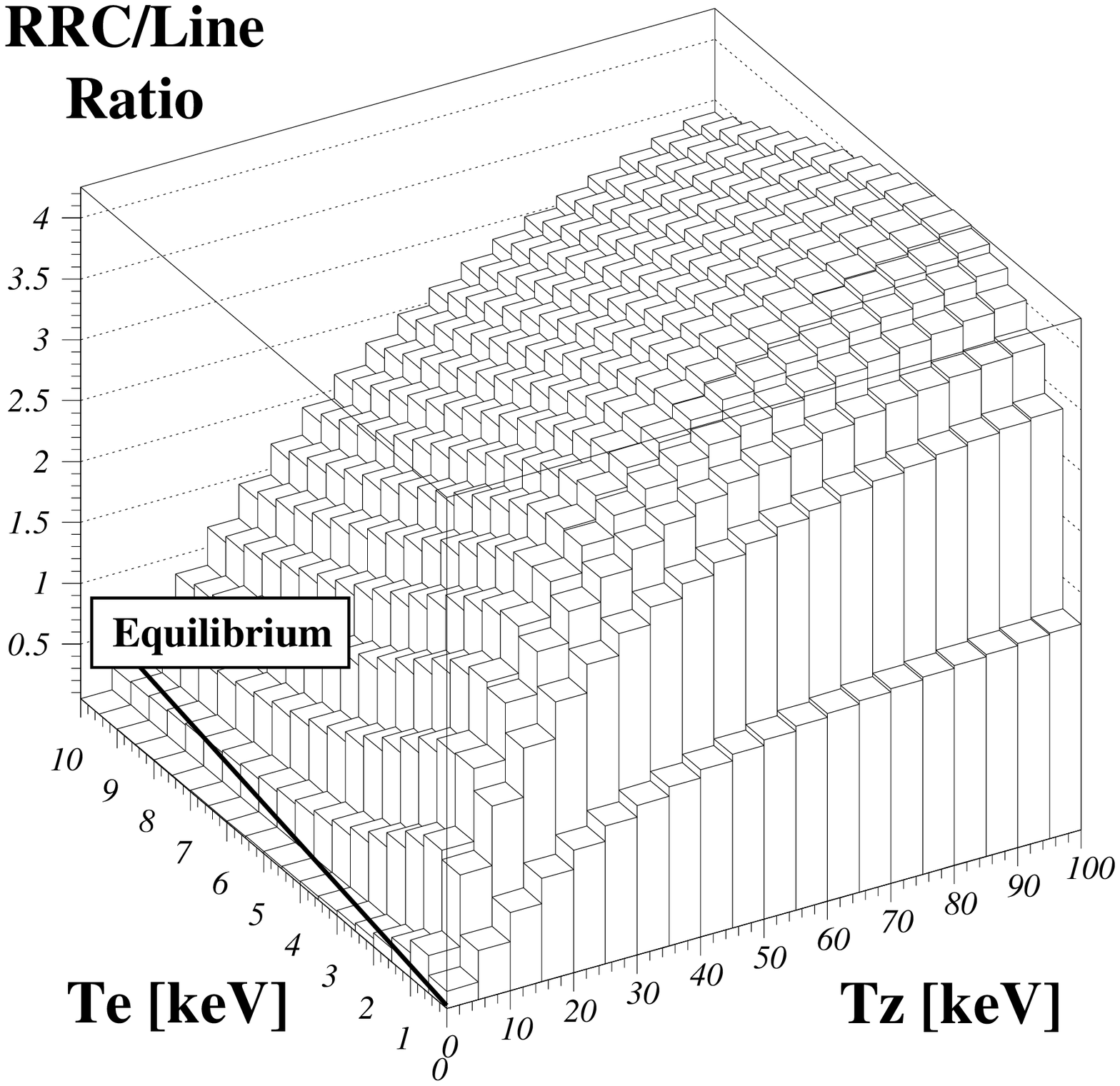}}} 
\end{center} 
\end{figure} 

\end{document}